\documentclass{aa}  

\usepackage{natbib}
\bibpunct{(}{)}{;}{a}{}{,} 

\usepackage{graphicx}
\usepackage{txfonts}
\begin{document}

\title{Light-curve analysis of KOI 2700b: the second extrasolar planet with a comet-like tail}
\author{Z. Garai}
\institute{Astronomical Institute, Slovak Academy of Sciences, 059 60 Tatranská Lomnica, Slovakia\\
\email{zgarai@ta3.sk}}
\date{Received ; accepted}

 
  \abstract
   {The \textit{Kepler} object KOI 2700b (KIC 8639908b) was discovered recently as the second exoplanet with a comet-like tail. It exhibits a distinctly asymmetric transit profile, likely indicative of the emission of dusty effluents and reminiscent of KIC 12557548b, the first exoplanet with a comet-like tail.}
   {The scientific goal of this work is to verify the disintegrating-planet scenario of KOI 2700b by modeling its light curve and to put constraints on various tail and planet properties, as was done in the case of KIC 12557548b.}
   {We obtained the phase-folded and binned transit light curve of KOI 2700b, which we subsequently iteratively modeled using the radiative-transfer code SHELLSPEC. We modeled the comet-like tail as part of a ring around the parent star and we also included the solid body of the planet in the model. During the modeling we applied selected species and dust particle sizes.}
   {We confirmed the disintegrating-planet scenario of KOI 2700b. Furthermore, via modeling, we derived some interesting features of KOI 2700b and its comet-like tail. It turns out that the orbital plane of the planet and its tail are not edge-on, but the orbital inclination angle is from the interval $[85.1,88.6]$ deg. In comparison with KIC 12557548b, KOI 2700b exhibits a relatively low dust density decreasing in its tail. We also derived the dust density at the beginning of the ring and the highest optical depth through the tail in front of the star, based on a tail-model with a cross-section of $0.05 \times 0.05 R_\odot$ at the beginning and $0.09 \times 0.09 R_\odot$ at its end. Our results show that the dimension of the planet is $R_\mathrm{p}/R_\mathrm{s}\leq 0.014$ ($R_\mathrm{p}\leq 0.871R_\oplus$, or $\leq 5551$ km). We also estimated the mass-loss rate from KOI 2700b, and we obtained $\dot{M}$ values from the interval $[5.05 \times 10^{7},4.41 \times 10^{15}]$ g.s$^{-1}$. On the other hand, we could not draw any satisfactory conclusions about the typical grain size in the dust tail.}
   {}


   \maketitle
%

\section{Introduction}
Strong irradiation in close-in exoplanet systems may cause mass loss from the planet \citep{Burrows1, Guillot1}, such as that detected, for example, in HD 209458b \citep{Vidal1, Vidal2} and in HD 189733b \citep{Lecavelier1, Bourrier1}. Several theoretical studies have been devoted to this subject \citep{Yelle1, Tian1, Hubbard1}. In certain cases the mass loss from the exoplanet may cause formation of a comet-like tail. The hypothesis that a close-in exoplanet may have a comet-like tail was suggested by \citet{Schneider1} and was revisited by \citet{Mura1}. The transit light curve of dusty extra-solar comets was investigated, for example, by \citet{Lamers1} and  \citet{Lecavelier2}. The first good evidence for exocomet transits was presented by \citet{Rappaport3}. 

The first exoplanet with a comet-like tail, KIC 12557548b, was discovered from \textit{Kepler} long-cadence data by \citet{Rappaport1}, and was found to be a close-in exoplanet with an extremely short orbital period of $P_\mathrm{orb} \simeq 0.65356$ days. \citet{Rappaport1} suggested that the planet's size is not larger than Mercury. The light curve of this planet was studied in more detail by \citet{Brogi1}, \citet{Budaj1}, \citet{Kawahara1}, \citet{Croll1}, \citet{Werkhoven1}, \citet{Bochinski1}, \citet{Schlawin1} and \citet{vanLieshout2}. \citet{Brogi1} and \citet{Budaj1} first validated the disintegrating-planet scenario using a model and both found that dust particles in the tail have typical radii of about 0.1-1 micron. \citet{Perez1} proposed a model of the atmospheric escape via the thermal wind that is only effective for planets which are less massive than Mercury. Gravity of the more massive planets would provide too deep a potential barrier for the wind. \citet{Garai1} searched for comet-like tails in a sample of 20 close-in exoplanet candidates with a period similar to KIC 12557548b from the \textit{Kepler} mission, however, none of the exoplanet candidates showed signs of a comet-like tail. This result is in agreement with the model proposed by \citet{Perez1}. 

Recently, two more exoplanets have been discovered, KOI 2700b and K2-22b, whose transit shapes show evidence of a comet-like tail \citep{Rappaport2, Sanchis1}. The \textit{Kepler} object KOI 2700b (KIC 8639908b) was discovered by \citet{Rappaport2} as the second exoplanet with a comet-like tail. It exhibits a distinctly asymmetric transit profile, likely indicative of the emission of dusty effluents and reminiscent of KIC 12557548b. The orbital period of KOI 2700b was determined by the discoverers using Lomb-Scargle analysis ($P_\mathrm{orb}=0.910023(4)$ days) and $\chi^2$ minimization ($P_\mathrm{orb}=0.910022(5)$ days). The host star has $T_\mathrm{eff} = 4435$ K, $M \simeq 0.63 M_{\odot}$ and $R \simeq 0.57 R_{\odot}$, comparable to the parameters ascribed to KIC 12557548. The upper limit of the planet radius $R_\mathrm{p} \sim 1.06 R_\oplus$, determined by the same authors, is in agreement with the theory of the thermal wind and planet evaporation \citep{Perez1}. The mass-loss rate, found by \citet{Rappaport2}, is $\sim 6 \times 10^9$ g.s$^{-1}$ or $\sim$ 2 lunar masses per Gyr. The composition of the dust ejected by KIC 12557548b and KOI 2700b was proposed by \citet{vanLieshout1}. The observed tail lengths are consistent with dust grains composed of corundum ($\alpha$-Al$_2$O$_3$) or iron-rich silicate minerals. Pure iron and carbonaceous compositions are not favored.                     

In this paper we aim to verify the disintegrating-planet scenario of KOI 2700b by modeling its light curve and put constraints on various tail and planet properties, as was done in the case of KIC 12557548b. The light curve of the object was analyzed similarly as in \citet{Budaj1}, however, in comparison with \citet{Budaj1}, in this work the light-curve modeling process is more sophisticated. We first improved the orbital period and constructed the phase-folded and binned transit light curve of KOI 2700b using long-cadence \textit{Kepler} observations from the quarters 1 -- 17 (Sect. \ref{constructingLC}). Subsequently, we searched for long-term orbital period variations (Sect. \ref{PDM}). We then calculated phase functions and used the radiative-transfer code SHELLSPEC \citep{Budaj4, Budaj2}. We applied this code iteratively to model the observed light curve of KOI 2700b. Mie absorption and scattering on spherical dust grains with realistic dust opacities, phase functions, and finite radius of the source of the scattered light were taken into account (Sect. \ref{modelingwithshell}). The resulting parameters and model light curves are described and discussed in Sect. \ref{res}. Our findings are concluded in Sect. \ref{cnc}.        

\section{The light curve of KOI 2700b}
\label{constructingLC}

\subsection{Observations}
\label{Obs}
\begin{figure}[t]
\centering
\includegraphics[width=80mm]{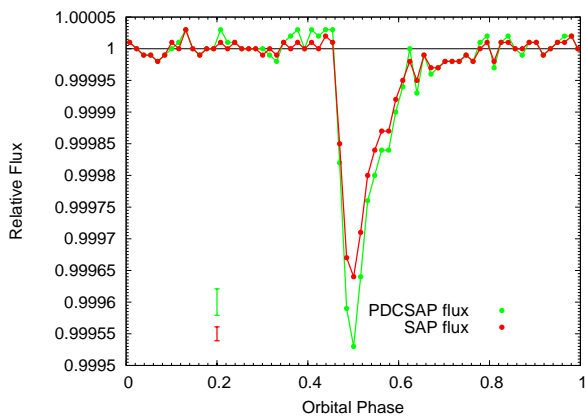}
\caption{Comparison between the SAP and PDCSAP \textit{Kepler} phase-folded and binned transit light curves of KOI 2700b. The illustrative vertical error bars on the left side represent the median uncertainties of data points depicted with the same color.}
\label{PDCSAPSAP}
\end{figure}

We used the publicly available \textit{Kepler} data from the quarters 1 -- 17 in the form of Simple Aperture Photometry (SAP) fluxes. Pre-search Data Conditioning Simple Aperture Photometry (PDCSAP) fluxes often deliver over-corrected light curves, in which the astrophysical signal is reduced or even canceled, hence we avoided using these data. A comparison between the SAP and PDCSAP \textit{Kepler} phase-folded and binned transit light curves of KOI 2700b is depicted in Fig \ref{PDCSAPSAP}. Only the long cadence data were used to construct the light curve. These are 64~842 observations with an exposure time of about 30 min. \textit{Kepler} observations were reduced in a similar manner to in \citet{Budaj1} and \citet{Garai1}. 

\begin{figure}[t]
\centering
\includegraphics[width=80mm]{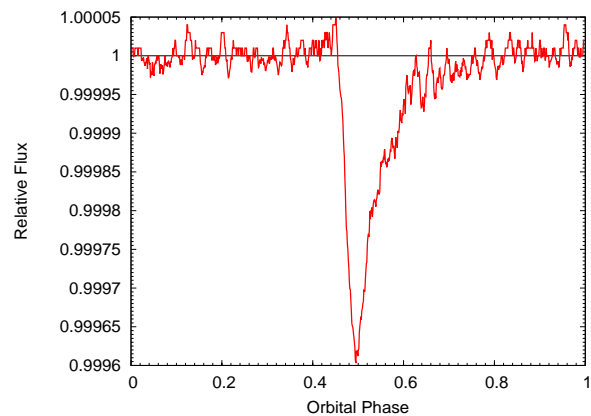}
\caption{Averaged light curve of KOI 2700b, smoothed using the running window technique. The window width and step was 0.011 and 0.0011, respectively (in units of phase). The light curve shows systematic fluctuations at short timescales (20 -- 30 min). See the text.}
\label{distrib_15min}
\end{figure}

\begin{figure}[t]
\centering
\includegraphics[width=80mm]{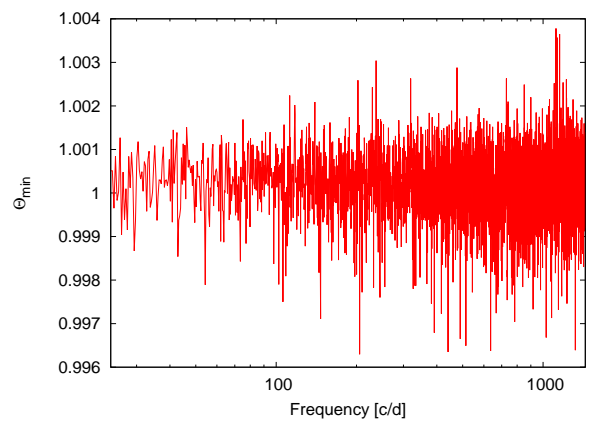}
\caption{Results of the period analysis of KOI 2700b. We searched for periodicities from 1 min to 1 hour using the method of phase dispersion minimization (see Sect. \ref{PDM}), but we did not find any significant frequency in this interval. See the text.}
\label{period}
\end{figure}

\begin{figure}[t]
\centering
\includegraphics[width=80mm]{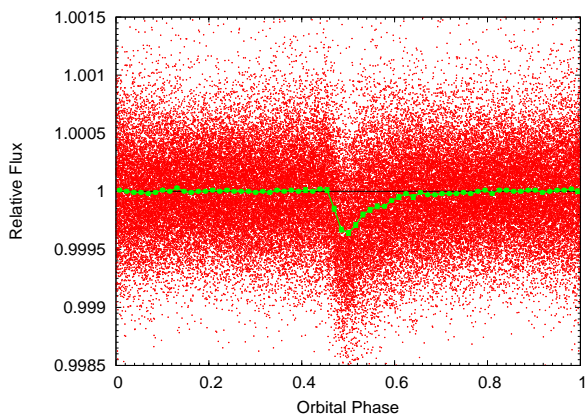}
\caption{Phase-folded and binned transit light curve of KOI 2700b. Red points are observations and green points represent the averaged data. We note that averaged data points have uncertainties that are  too small to be discerned in this plot. Error bars of these data are well depicted, for example, in Fig \ref{alumina_enstatite_a2m10}.
}
\label{KOI2700LC}
\end{figure}

Each quarter has a different flux level. Consequently, fluxes within each quarter were normalized to unity. We then improved the orbital period of the exoplanet found by \citet{Rappaport2}. For this purpose we used the method of phase dispersion minimization described in Sect. \ref{PDM}. First we phased the data with the orbital period of $P_\mathrm{orb}=0.910023$ days \citep{Rappaport2}. We used the phase 0.5 for transits. Subsequently, data were cut into segments each covering one orbital period. Each segment of data was fitted with a linear function. During the fitting procedure the part of the data between phases 0.45 and 0.65, covering the transit, was excluded from the fit. Consequently, the linear trend was removed from each chunk of data (including the transit data). This method can effectively remove the long term variability (mainly variability of the host star due to spots and rotation) while it does not introduce any nonlinear trend to the phased light curve. The final value of the orbital period was then found in these detrended data. We found the orbital period of $P_\mathrm{orb}=0.9100259(15)$ days via the phase dispersion minimization method, which is in good agreement with the period presented by the discoverers. Finally, the data were phased with this new orbital period. Subsequently, we checked our result using the Fourier method \citep{Deeming1} and found $P_\mathrm{orb}=0.910042(7)$ days. This method confirmed the orbital period resulting from the phase dispersion minimization method. We estimated the error by performing Monte Carlo simulations generating and analyzing 100 artificial datasets. 

To reduce the noise, the phased light curve was subjected to an averaging. We first applied the running window averaging technique. We used a window with the width in units of phase of 0.00075 (1-min width, negligible in comparison with the exposure time), 0.005 (6.5-min width, according to \citet{Rappaport2}), 0.011 (15-min width, about 1/2 of the exposure time) and 0.023 (30-min width, comparable with the exposure time). The step was set as 1/10 window width in every case. Since the obtained averaged light curves showed systematic fluctuations at short timescales (about 20 -- 30 min), easily visible, for example, in the light curve smoothed using the 15-min-width window (Fig. \ref{distrib_15min}), we performed period analysis with the main aim to explore its nature, using the method of phase dispersion minimization (see Sect. \ref{PDM}). For this purpose we used the available short cadence SAP \textit{Kepler} detrended data, and searched for periodicities from 1 min to 1 hour. Figure \ref{period} shows that there are no significant frequencies in this interval, indicating that the observed systematic fluctuations of averaged light curves may be due to poor SNR or to the exposure time. Subsequently, we applied averaging via data binning. Instead of stable window width, this technique uses a certain number of data points per bin. We examined the cases from 500 to 1500 data points per bin with a step of 50 data points per bin, and finally we selected the case of 1000 data points per bin, according to the minimum scatter and relatively good coverage of the phased light curve with averaged data points. The obtained light curve is depicted in Fig. \ref{KOI2700LC}, as well as in Fig. \ref{PDCSAPSAP} (SAP flux). No systematic fluctuations are visible in this case and, if not mentioned otherwise, we used this averaged light curve in our analysis.         

\subsection{Comparison with the light curve of KIC 12557548b}
\begin{figure}
\centering
\includegraphics[width=80mm]{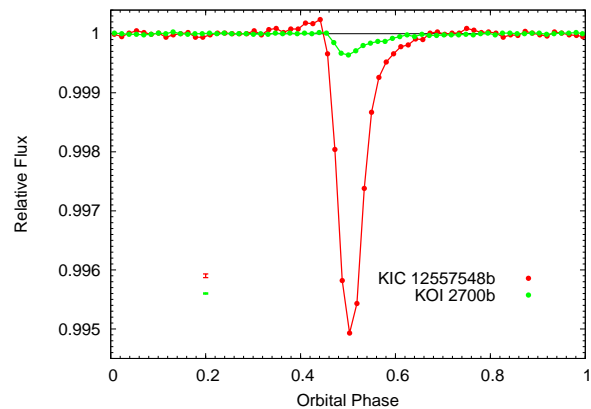}
\caption{Comparison between the SAP \textit{Kepler} phase-folded and binned transit light curve of exoplanets KIC 12557548b and KOI 2700b. The illustrative vertical error bars on the left side represent the median uncertainties of data points depicted with the same color.}
\label{KICKOI}
\end{figure}

The light curve of the exoplanet KOI 2700b is peculiar and very interesting. It exhibits a distinctly asymmetric transit profile with a sharp ingress followed by a short sharp egress and a long smooth egress; reminiscent of the light curve of the exoplanet KIC 12557548b. The observed light curve of the exoplanet KOI 2700b might show a relatively weak pre-transit brightening (see for example, Fig. \ref{PDCSAPSAP}), however, we could not statistically identify this light-curve feature, previously found at KIC 12557548b as significant. This latter exoplanet also exhibits a relatively weak post-transit brightening (Fig. \ref{KICKOI}). These brightenings are caused by the forward scattering on dust particles in the tail \citep{Brogi1, Budaj1}. The light curves have one more similarity. The transit ingress begins at an orbital phase of about 0.45 and the transit egress ends at an orbital phase of about 0.65 (Fig. \ref{KICKOI}). The transit duration in phase units is 0.2 in both cases. The transit depth, however, is very different. KIC 12557548b has an average transit depth of about 0.005 in flux (i.e., 0.5\%), while KOI 2700b has about ten times shallower average transit depth, 0.0004 in flux (i.e., 0.04\%).   

\begin{figure}
\centering
\includegraphics[width=80mm]{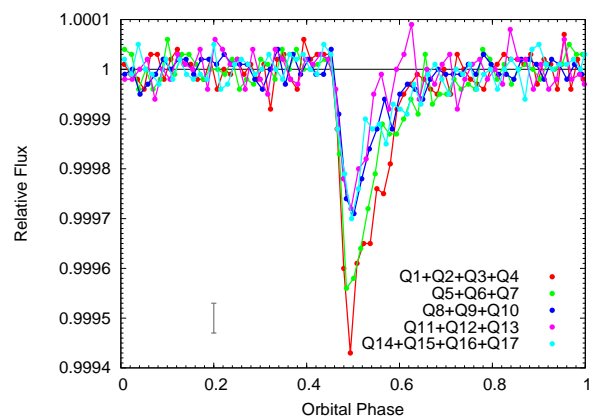}
\caption{Transit depth decreasing at KOI 2700b. The phase-folded and binned transit light curves of the exoplanet were constructed from 3 and 4 quarters (Q) of detrended data. The illustrative vertical error bar on the left side represents the median uncertainty of all data points.}
\label{KOI_divide}
\end{figure}

\subsection{Evolution of the light curve}
The relatively shallow average transit depth of the exoplanet KOI 2700b ($\sim$ 0.04\%) makes it difficult to study possible transit-to-transit variations in the transit depth, as was done in the case of KIC 12557548b. On the other hand, \citet{Rappaport2} reported that the transit depth appears to decrease systematically with time during the \textit{Kepler} mission. During this step we first divided our detrended data into five approximately equal segments (in time) and each segment of data was subsequently phased with the orbital period of $P_\mathrm{orb}=0.9100259$ days (Sect. \ref{Obs}), found via the method of phase dispersion minimization (see Sect. \ref{PDM}). Since we used less data points per light curve (approximately 13~000 vs. 64~000), we applied 200 data points per bin (instead of 1000 data points per bin) during the data averaging. The resulting phased light curves are depicted in Fig. \ref{KOI_divide}. Based on our analysis we can confirm the result concerning
decreasing  transit depth, found by \citet{Rappaport2}. On the other hand, we can also conclude that this depth decreasing is not monotonical. The last three light curves have approximately the same transit depth, which may indicate the possibility of a long-term periodic variability as well.

\section{Search for long-term orbital period variations}
\label{PDM}
\begin{figure}
\centering
\includegraphics[width=80mm]{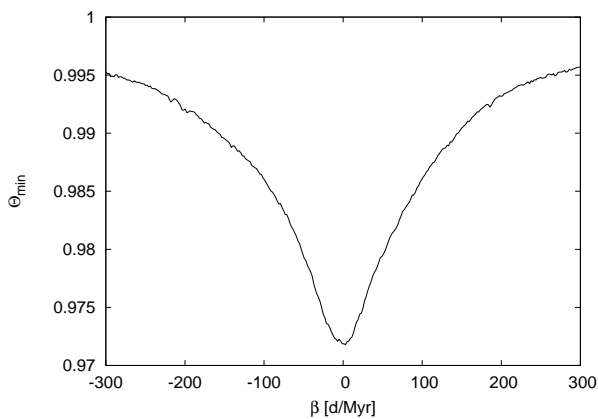}
\caption{Search for a long-term orbital period change of KOI 2700b. The minimum of the parameter $\Theta_\mathrm{min}$ corresponds to the period-change speed $\beta = 1.3 \pm 3.2$ d/Myr, which means that there is no significant evidence for the long-term orbital period change.}
\label{KOI2700_PDM}
\end{figure}

\citet{Perez1} suggest that disintegrating planets undergo negligible orbital period variations due to the evaporation process. During this step we examined this prediction via searching for possible long-term changes of the orbital period of KOI 2700b. For this purpose we used the method of phase dispersion minimization \citep{Stellingwerf1}, application PDM2, version 4.13\footnote{Actual version of the software package is available on the web-page \url{http://www.stellingwerf.com}.}, which uses bin structure 50/2. The detrended data were used for this purpose (see Sect. \ref{Obs}). We assumed that the period changes linearly:

\begin{equation}
  \label{equation1}
  P=P_{\mathrm{0}}+\beta t
,\end{equation}

\noindent where $\beta$ is a dimensionless value, but is often expressed in days/million years (d/Myr). The output from the analysis is a curve (Fig. \ref{KOI2700_PDM}), which shows the dependence of $\Theta_\mathrm{min}$ as a function of $\beta$. The term $\Theta_\mathrm{min}$ is a dimensionless statistical parameter \citep{Stellingwerf1}. The minimum value of $\Theta_\mathrm{min}$ indicates the speed of the period change -- $\beta$. We fitted a parabola to this curve and obtained $\beta = 1.3 \pm 3.2$ d/Myr, which means that there is no significant evidence for the long-term orbital period change during the time span of the \textit{Kepler} observations. This result is in agreement with the prediction, presented by \citet{Perez1}. The error was estimated by means of Monte Carlo simulations. We generated and analyzed 100 artificial datasets with the same standard deviation as the original data.

\section{Light-curve modeling with the code SHELLSPEC}
\label{modelingwithshell}
\subsection{Calculation of the optical properties of the dust}

According to \citet{vanLieshout1}, the comet-like tail of KOI 2700b may consist of dust grains composed of corundum ($\alpha$-Al$_2$O$_3$) or iron-rich silicate minerals. Therefore during the modeling we applied selected species, which have similar chemical composition: $\gamma$-alumina, enstatite, forsterite, olivine (with 50\% magnesium and 50\% iron) and pyroxene (with 40\% magnesium and 60\% iron). Aluminum oxide (Al$_2$O$_3$) occurs naturally in its crystalline polymorphic phase as the mineral corundum, which is $\alpha$-alumina. Aluminum oxide also exists in many other phases; each has a unique crystal structure and properties. In our model we applied cubic $\gamma$-Al$_2$O$_3$. Enstatite is the magnesium-rich end-member of the pyroxene silicate mineral series enstatite (MgSiO$_3$) -- ferrosilite (FeSiO$_3$). Forsterite (Mg$_2$SiO$_4$) is the magnesium-rich end-member of the olivine solid solution series. It is isomorphous with the iron-rich end-member, fayalite. Besides forsterite, in our model we also applied olivine with 50\% magnesium and 50\% iron. The pyroxenes are a group of minerals with the general formula XY(Si,Al)$_2$O$_6$. The group has 20 end-members and one of the end-members is enstatite (see above). Pyroxenes are, however, usually isomorphous mixtures of some end-members. Therefore, in our model we also applied a pyroxene with 40\% magnesium and 60\% iron.          

Calculating the optical properties (i.e., opacities and phase functions, see below) of the dust grains composed of a mineral is relatively difficult and requires a lot of computing time. That is why \citet{Budaj3} prepared tables of phase functions, opacities, albedos, equilibrium temperatures and radiative accelerations of dust grains in exoplanet systems. The tables cover the wavelength range of 0.2 to 500 microns and 21 particle radii from 0.01 to 100 microns for several species. Their assumptions include spherical grain shape, Deirmendjian particle size distribution \citep{Deirmendjian1} and Mie theory. \citet{Budaj3} used a widely available Mie scattering code CALLBHMIE\footnote{The code is available on the web-page \url{https://svn.ssec.wisc.edu/repos/geoffc/Mie/bhmie-f/callbhmie.f}.}, which calls iteratively the Mie scattering subroutine BHMIE \citep{Bohren1}. The tables are freely available on the web\footnote{See the web-page \url{https://www.ta3.sk/~budaj/dust/deirm/}.} and are applicable for our purposes in the code SHELLSPEC. From these tables we selected opacities and phase functions for $\gamma$-alumina, enstatite, forsterite, olivine (with 50\% magnesium and 50\% iron) and pyroxene (with 40\% magnesium and 60\% iron). In our model we applied the following particle radii: 0.01, 0.1 and 1.0 micron.

\begin{figure}
\centering
\includegraphics[width=80mm]{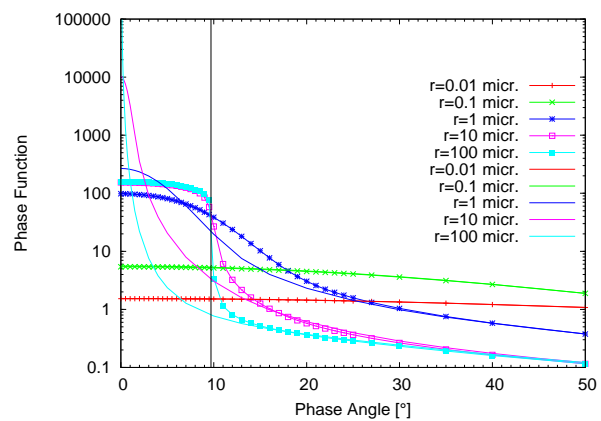}
\caption{Phase functions at 0.6 micron for different dust particle radii of $\gamma$-alumina. Lines without symbols are phase functions assuming a point source of light. Lines with symbols are phase functions that take into account the finite dimension of the stellar disk. The vertical line illustrates the angular dimension of the stellar disk as seen from KOI 2700b.}
\label{phasef}
\end{figure}

Dust can absorb the impinging radiation and convert it directly into heating of the grains. This process is called \textit{absorption} or \textit{true absorption} and it is quantified by the \textit{absorption opacity}. Dust can also scatter radiation in a process called \textit{scattering} without being heated. This process is quantified by the \textit{scattering opacity}. The sum of the absorption opacity and the scattering opacity is the \textit{total opacity} (or simply the \textit{opacity}). Furthermore, scattering can be highly asymmetric, a property that is described by means of the phase function, which depends on the scattering angle (the deflection angle from the original direction of the impinging radiation). The most prominent feature is a strong forward scattering, when the scattering angle is nearly zero. Brightenings on the light curve of KIC 12557548b are also caused by the forward scattering on dust particles in the tail \citep{Brogi1, Budaj1}. We also have to take into account the fact that from the viewpoint of dust particles the parent star has a non-negligible angular dimension on the sky. Dust particles of KOI 2700b in the distance of $a = 3.363R_{\odot}$ from the parent star view this star as a disk with an angular diameter of about $19.4$ deg. The true dimension of this star is $R_\mathrm{s} = 0.57 R_{\odot}$ \citep{Rappaport2}. To take this effect into account we have to split the stellar disk into elementary surfaces and integrate the phase function over the disk. We calculated phase functions with a very fine step in the interval $[0,10]$ deg, because of the strong forward scattering and, consequently, the disk averaged phase function with a very fine step near the edge of the stellar disk. For this purpose we applied the software DISKAVER\footnote{The software is available on the web-page \url{https://www.ta3.sk/~budaj/dust/deirm/diskaver/}.}. It assumes a quadratic limb darkening of the stellar surface: 

\begin{equation}
\label{qadraticlimbdarklaw}
I_\nu = I_\nu(0)[1-u_1(1 - \cos \theta) - u_2(1 - \cos \theta)^2],
\end{equation}  
 
\noindent where $I_\nu(0)$ is intensity perpendicular to the surface of the source and $\theta$ is angle between the line of sight and a normal to the surface. The quadratic limb darkening coefficients $u_1$ and $u_2$ were linearly interpolated based on the stellar parameters $T_\mathrm{eff} = 4433$ K, $\log g = 4.721$ (cgs) and $\mathrm{Fe/H}=-0.2$ \citep{Rappaport2} for the \textit{Kepler} passband, using the on-line applet EXOFAST -- Quadratic Limb Darkening\footnote{See the web-page \url{http://astroutils.astronomy.ohio-state.edu/exofast/limbdark.shtml}.}, which is based on the IDL-routine QUADLD \citep{Eastman1}, as $u_1 = 0.622$ and $u_2 = 0.113$. This software interpolates the \citet{Claret1} quadratic limb darkening tables. Calculations with the software DISKAVER were performed at a wavelength of 0.6 microns for consistency with the \textit{Kepler} passband. An illustration of such phase functions,
which take into account the finite dimension of the source of light, is depicted in the Fig. \ref{phasef}. Corresponding phase functions without taking into account the finite dimension of the source of light are also shown. These exhibit a strong peak near phase angle zero, which is the so-called forward-scattering peak. Larger particles and/or shorter wavelengths tend to have a stronger and narrower forward-scattering peak than smaller particles and/or longer wavelengths. Therefore, taking into account the finite dimension of the source of light is important mainly for larger particles and/or shorter wavelengths. 
 
\subsection{Model and light-curve modeling} 
\label{modeling}

The phased and averaged transit light curve of the exoplanet KOI 2700b (Fig. \ref{KOI2700LC}) was in principle modeled using the radiative-transfer code SHELLSPEC\footnote{The software is available on the web-page \url{https://www.ta3.sk/~budaj/shellspec.html}.} \citep{Budaj4, Budaj2}, version 39. This code calculates the light curves and spectra of interacting binaries or exoplanets immersed in the three-dimensional (3D) circum-stellar or circum-planetary environment. It solves simple radiative transfer along the line of sight and the scattered light is taken into account under the assumption that the medium is optically thin. A number of optional objects (such as a spot, disk, stream, ring, jet, shell) can be defined within the model, or it is possible to load a precalculated model from an extra file. Synthetic light curves or trailing spectrograms can be produced by changing our viewpoints on the 3D object. The software is written in Fortran77, however, some previous versions were transformed into the language Fortran90; see for example, \citet{Tkachenko1, Sejnova1}.
  
For our purpose we used an optional object in the form of a ring, which we subsequently modified. From the transit shape we can conclude that transits are caused primarily by the comet-like tail, and not by the solid body of the planet. We can model such a comet-like tail as part of a ring with a non-negligible thickness around a central star. Therefore during our calculations we assumed a spherical and limb-darkened central star (using the coefficients $u_1 = 0.622$ and $u_2 = 0.113$ from Eq. (\ref{qadraticlimbdarklaw})), with a radius of $R_\mathrm{s} = 0.57 R_{\odot}$, mass of $M_\mathrm{s} = 0.632 M_\odot$ and effective temperature of $T_\mathrm{eff} = 4433$ K \citep{Rappaport2}, located in the geometrical center of our ring. We modeled the comet-like tail as part of a ring with a radius of $a = 3.363R_{\odot}$. Its geometrical cross-section is monotonically enlarging from the planet to the end of the ring, which is located at 60 deg behind the planet. At this point the ring is truncated. In our calculations we applied the same length of the tail as per \citet{Budaj1}, because the transit duration in phase units is very similar in both cases (Fig. \ref{KICKOI}). The cross-section of the ring \textit{C} and dust density along the ring $\rho$ are allowed to change with the angle \textit{t} [rad]:

\begin{equation}
\label{densitychangeA1}
\rho(t) = \rho(0)\frac{C(0)}{C(t)}(|t-t(0)|/\pi+1)^{A1},
\end{equation}

\noindent or

\begin{equation}
\label{densitychangeA2}
\rho(t) = \rho(0)\frac{C(0)}{C(t)}e^{(|t-t(0)|A2)/\pi},
\end{equation}

\noindent where $\rho(0)$, $C(0),$ and $t(0)$ are the dust density, cross-section, and phase angle of view at the beginning of the ring and $A1$, $A2$ are the density exponents to model the dust destruction in the tail. We can see that there is a strong degeneracy between the cross-section and the dust density at a certain phase angle. In general, if we increase the cross-section and we want to obtain an appropriate model of the observed light curve, the result is that we need to decrease the dust density. There is a degeneracy relation $C\rho=$ const. Since the dimension of the dust tail of KOI 2700b is unknown, we arbitrarily defined a geometrical cross-section of the tail, which was preferable for our computation process (in terms of the grid density, grid dimension, and computing time). Therefore, in our calculations we assumed a dust tail with a cross-section of $0.05 \times 0.05 R_\odot$ at the beginning and $0.09 \times 0.09 R_\odot$ at its end. These values are also in agreement with the escape velocity from a Mercury-sized small planet (a few km.s$^{-1}$). We note that in certain cases the chosen cross-section of the tail can affect the results about the grain size; for example, at higher inclinations ($i<78$ deg) large amounts of dust will not transit the parent star and this can affect the identification of the typical particle radii of the tail (described in Sect. \ref{dustres}). However, as we described in Sect. \ref{tailres}, we obtained $i$ in the range $[85.1,88.6]$ deg (the average value is $i=87.06$ deg), that is, the latter situation is not the case here. The ring was located in the orbital plane of the planet. 

On the other hand, based on the transit depth, we can assume that there is still ample room for a contribution to the light curve from the solid body of the planet. Therefore, we also included the solid body of the planet in the model, defined as a dark, non-transparent sphere and located at the beginning of the ring. The dimension of the planet was parametrized as $R_\mathrm{p}/R_\mathrm{s}$ (ratio of the radii). Although \citet{Rappaport2} present the formal upper limit of the planet radius of $R_\mathrm{p} \leq 1.06R_{\oplus}$ ($2\sigma$ limit), which roughly corresponds to $R_\mathrm{p}/R_\mathrm{s} \leq 0.017$, based on the theory of the thermal wind and planet evaporation \citep{Perez1} we can expect that the planet itself must be Mercury-sized, or smaller than Mercury, which corresponds to $R_\mathrm{p}/R_\mathrm{s}\leq0.006$. 

In the SHELLSPEC code the central star with the defined object is located in a 3D grid. The code enables the user to look on the grid from different points of view and to calculate the corresponding flux. The flux is always calculated in the observer’s line of sight. The orbital inclination angle $i$ corresponds the inclination of the intrinsic rotation axis of the model to the line of sight. At each point of view we calculated the final flux as $(s+r)/s = f$, where $s$ means modeling the flux from the parent star, $s+r$ means modeling the parent star with the ring and planet, and $f$ is the final and normalized flux from the system. In this way we also eliminated fluctuations due to the grid structure of objects in our model. To speed up the computation process, we reduced the number of points by selecting the near-transit part of the light curve. We calculated 33 artificial data points per synthetic light curve, between phases 0.25 and 0.75, in the same phases as per the observed light curve. This cropping of the synthetic light curve is justified, since we are interested only in the near-transit region. The synthetic light curves were subsequently convolved with a box-car with a width of 30 minutes, simulating the integration time of the \textit{Kepler} long cadence exposure. Convolved light curves were used for comparison with the observed light curve. 

For the modeling process we generated and used an iterative procedure, which applied the SHELLSPEC code as a subroutine, and searched for the best fit parameters. Four free parameters were adjusted simultaneously during the fitting procedure -- the orbital inclination angle $i$ [deg], the dust density at the beginning of the ring $\rho(0)$ [g.cm$^{-3}$], the ratio of the radii $R_\mathrm{p}/R_\mathrm{s}$, and the density exponent $A2$. Since model light curves calculated using $A1$ and $A2$ differed only slightly, we decided to use the density exponent $A2$ during the modeling, because it explains the dust destruction in the tail better than $A1$. One more free parameter -- the transit midpoint phase shift of the synthetic light curve with respect to the observed light curve ($\Delta\varphi_0$) -- was adjusted only before the modeling process and then was kept fixed to its best value. This parameter reflects the unknown mid-transit time of the planet. Every synthetic light curve was shifted in phase by $\Delta\varphi_0=-0.235$. The advantage of this treatment is that it saves computing time; on the other hand we cannot exclude the possibility that uncertainties on the resulting parameters are underestimated. We prepared a parameter space with a certain stepping, composed from the above-mentioned four free parameters and limited based on the previous step-wise test calculations as follows. The parameter $i$: $[90,80]$ deg; the parameter $\rho(0)$: $[0.5000,0.0001] \times 10^{-15}$ g.cm$^{-3}$; the parameter $R_\mathrm{p}/R_\mathrm{s}$: $[0.017,0.001]$; and the parameter $A2$: $[-1.0,-25.0]$. During the fitting procedure all combinations of free parameters were examined and the observed light curve was compared with the corresponding model light curve. A formal, quantitative goodness-of-fit was measured via determination of reduced $\chi^2$ ($\chi^2_{\mathrm{red}}$). The best fit corresponds to the minimum value of $\chi^2_{\mathrm{red}}$. During the next iteration we reduced the range of the parameters and at the same time we used a finer stepping in the parameter space to fit the observed transit light curve better. As the first iteration we used the above-mentioned range of the parameters and the stepping as follows. The parameter $i$: 2.0 deg; the parameter $\rho(0)$: 0.0100 $\times 10^{-15}$ g.cm$^{-3}$; the parameter $R_\mathrm{p}/R_\mathrm{s}$: 0.004; and the parameter $A2$: 4.0. We selected the best value of a given parameter, found in the previous iteration, as a median ($\tilde{x}_\mathrm{par}$) value of the new parameter range. As the second iteration we used the following range/stepping of the parameters: $\tilde{x}_i$ $\pm$ 2.0, stepping 1.0 deg; $\tilde{x}_{\rho(0)}$ $\pm$ 0.0100 $\times 10^{-15}$, stepping 0.0010 $\times 10^{-15}$ g.cm$^{-3}$; $\tilde{x}_{R_\mathrm{p}/R_\mathrm{s}}$ $\pm$ 0.004, stepping 0.002; and $\tilde{x}_{A2}$ $\pm$ 4.0, stepping 1.0 and during the last (third) iteration, similarly: $\tilde{x}_i$ $\pm$ 1.0, stepping 1.0 deg; $\tilde{x}_{\rho(0)}$ $\pm$ 0.0010 $\times 10^{-15}$, stepping 0.0001 $\times 10^{-15}$ g.cm$^{-3}$; $\tilde{x}_{R_\mathrm{p}/R_\mathrm{s}}$ $\pm$ 0.002, stepping 0.001; and $\tilde{x}_{A2}$ $\pm$ 1.0, stepping 0.5. To obtain final results we applied these three iterations. In this way we reduced the computing time. 

To derive uncertainties on the resulting parameters we applied Monte Carlo simulations. First, we generated an artificial dataset with the same standard deviation as the observed data. Then, the artificial data were fitted with a model light curve and the best fit parameters were determined. The range of the parameters and the stepping corresponded to the range and stepping used in the last (i.e., in the 3rd) iteration (see the end of the previous paragraph). We note that the full procedure is not needed in this case, since the artificial data differ only slightly from the observed data. This process was repeated 100 times and, subsequently, to estimate the uncertainty in the given parameter the standard deviation was calculated. 

We executed 15 joint fitting procedures -- one for each combination of species and dust particle size. When describing our main results in Sect. \ref{res} we use the average value (without uncertainties) and the obtained range of the given parameter (including uncertainties). For individual values of parameters we refer the reader to Table \ref{partab}.

\begin{table*}
\caption{Overview of the resulting parameter values.}
\centering
\begin{tabular}{lcccccc}
\hline
\hline
Species                 & $i$ [deg]             & $A2$                 & $^{5}$$\rho(0)$ ($\times 10^{-15}$) [g.cm$^{-3}$]        & $^{5,6}$$\tau_\mathrm{max}$ ($\times 10^{-3}$)       & $R_\mathrm{p}/R_\mathrm{s}$   & $\chi^2_{\mathrm{red}}$       \\
\hline
\multicolumn{7}{c}{$r=0.01$ micron}                                                                                                                                                      \\
$\gamma$-Alumina        & $86\pm0.9$            & $-7.0\pm0.2$         & $0.2725\pm0.0003$               & $4.37\pm0.07$         & $0.009\pm0.003$               & 1.072                   \\
Enstatite               & $87\pm0.7$            & $-7.0\pm0.2$         & $0.2980\pm0.0009$               & $4.41\pm0.06$         & $0.008\pm0.003$               & 1.082                   \\
Forsterite              & $87\pm0.8$            & $-7.0\pm0.4$         & $0.2800\pm0.0006$               & $4.42\pm0.06$         & $0.008\pm0.002$               & 1.079                   \\
Olivine$^{1}$           & $88\pm0.4$            & $-7.0\pm0.3$         & $0.1600\pm0.0003$               & $4.36\pm0.05$         & $0.009\pm0.002$               & 1.075                   \\
Pyroxene$^{2}$          & $87\pm0.7$            & $-7.0\pm0.5$         & $0.3325\pm0.0008$               & $4.36\pm0.08$         & $0.009\pm0.003$               & 1.075                   \\
\hline
\multicolumn{7}{c}{$r=0.1$ micron}                                                                                                                                                      \\                                                                                               
$\gamma$-Alumina        & $88\pm0.4$            & $-7.0\pm0.3$         & $0.0222\pm0.0004$               & $4.55\pm0.08$         & $0.009\pm0.005$               & 1.113                   \\
Enstatite               & $87\pm0.9$            & $-7.0\pm0.2$         & $0.0252\pm0.0002$               & $4.59\pm0.06$         & $0.009\pm0.004$               & 1.131                   \\
Forsterite              & $87\pm0.2$            & $-7.0\pm0.3$         & $0.0220\pm0.0005$               & $4.57\pm0.07$         & $0.009\pm0.003$               & 1.137                   \\
Olivine$^{1}$           & $87\pm0.5$            & $-7.0\pm0.3$         & $0.0155\pm0.0004$               & $4.52\pm0.05$         & $0.009\pm0.004$               & 1.100                   \\
Pyroxene$^{2}$          & $87\pm1.0$            & $-7.0\pm0.4$         & $0.0174\pm0.0005$               & $4.55\pm0.06$         & $0.009\pm0.003$               & 1.113                   \\
\hline
\multicolumn{7}{c}{$r=1$ micron}                                                                                                                                                         \\
$\gamma$-Alumina        & $87\pm0.6$            & $-8.0\pm0.2$         & $0.2190\pm0.0005$               & $7.23\pm0.03$         & $0.007\pm0.006$               & 1.093                   \\
Enstatite               & $87\pm0.8$            & $-7.5\pm0.3$         & $0.2300\pm0.0006$               & $7.36\pm0.07$         & $0.007\pm0.007$               & 1.158                   \\
Forsterite              & $88\pm0.6$            & $-7.5\pm0.6$         & $0.2275\pm0.0004$               & $7.28\pm0.07$         & $0.007\pm0.005$               & 1.165                   \\
Olivine$^{1}$           & $87\pm0.5$            & $-7.0\pm0.4$         & $0.1890\pm0.0007$               & $6.01\pm0.04$         & $0.009\pm0.003$               & 1.106                   \\
Pyroxene$^{2}$          & $86\pm0.9$            & $-7.0\pm0.3$         & $0.1900\pm0.0006$               & $6.06\pm0.09$         & $0.009\pm0.004$               & 1.110                   \\
\hline
Average values$^{3}$    & $87.06$               & $-7.13$              & $0.1667$                        & $5.24$                & $0.008$                       & 1.107                   \\
Extreme values$^{4}$    & $85.1,88.6$           & $-6.5,-8.2$          & $0.0151,0.3333$                 & $4.28,7.43$           & $0.0,0.014$                    & $1.072,1.165$         \\                               
\hline
\hline                                          
\end{tabular}
\tablefoot{We executed 15 joint fitting procedures -- one for each combination of species and dust particle size. The table also contains the quantitative goodness-of-fit, parametrized as reduced $\chi^2$ ($\chi^2_{\mathrm{red}}$). $^{1}$With 50\% magnesium and 50\% iron. $^{2}$With 40\% magnesium and 60\% iron. $^{3}$Without uncertainties. $^{4}$Including uncertainties. $^{5}$These values were derived based on a tail-model with a cross-section of $0.05 \times 0.05 R_\odot$ at the beginning and $0.09 \times 0.09 R_\odot$ at its end (see Sect. \ref{modeling}). $^{6}$$\tau_\mathrm{max}$ is a derived parameter, which follows from $\rho(0)$.}     
\label{partab}
\end{table*}     

\section{Results and Discussion}
\label{res}
\subsection{Inclination of the tail}
\label{tailres}
The orbital inclination angle $i$ is a very important system parameter. We first assumed that the orbital plane of the exoplanet and its tail have an inclination of $i=90$ deg with respect to the plane of the sky. Our expectations, however, were not confirmed. During the modeling process it turned out that the transit of KOI 2700b is not edge-on, but based on our models, the orbital inclination angle is very probably close to the value of $i=87$ deg. We obtained $i$ in the range  $[85.1,88.6]$ deg\footnote{Converting the value of $i=87$ deg to the transit impact parameter gives $b=0.308$ and the range of values $[85.1,88.6]$ deg corresponds to $[0.504,0.144]$.} (see Table \ref{partab} for individual values). We obtained the value of $i\approx88$ deg in the cases of 0.1-micron $\gamma$-alumina grains, 1-micron forsterite grains and 0.01-micron olivine grains. Similarly, we obtained a value of $i\approx86$ deg in the cases of 0.01-micron $\gamma$-alumina grains and 1-micron pyroxene grains. In other cases we obtained for this parameter the value of $i\approx87$ deg. The resulting parameter values are, however, in agreement within a $1\sigma$ limit (see Table \ref{partab}).

\begin{figure*}
\centering
\centerline{
\includegraphics[width=80mm]{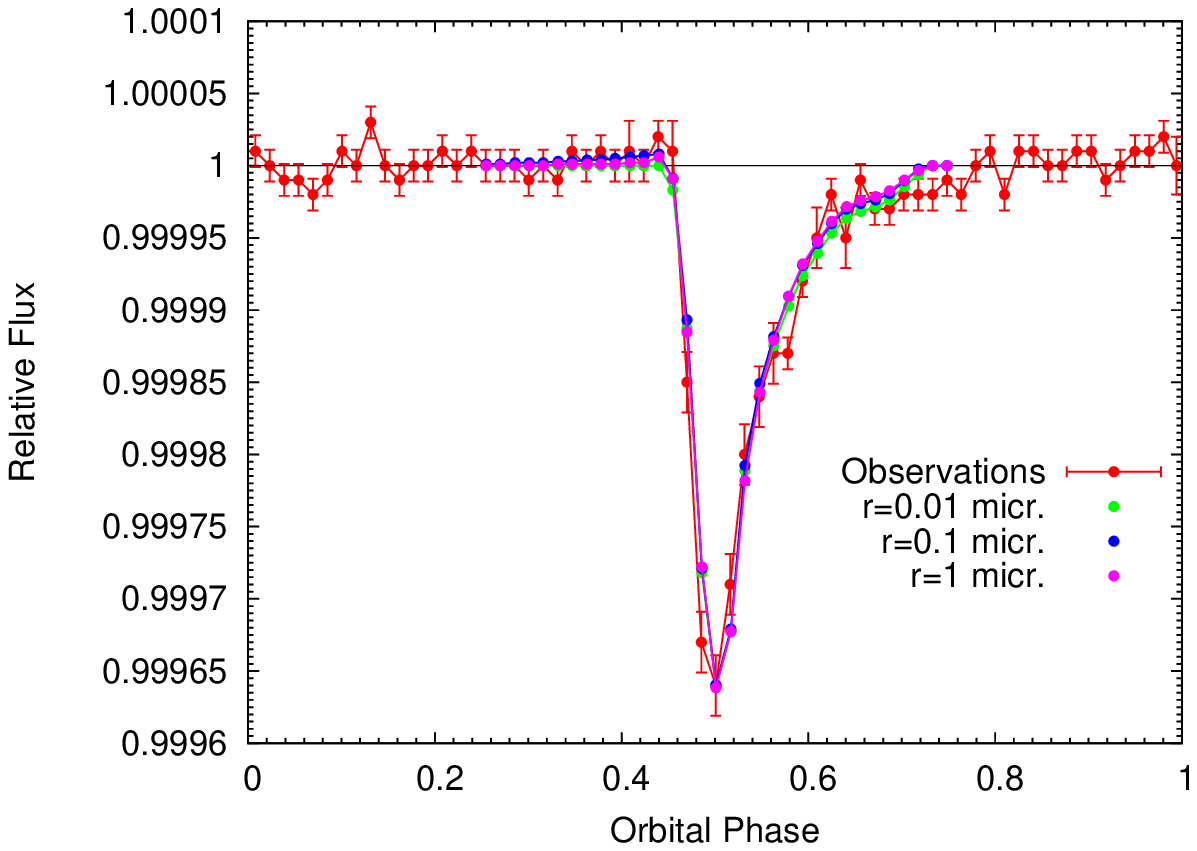}
\includegraphics[width=80mm]{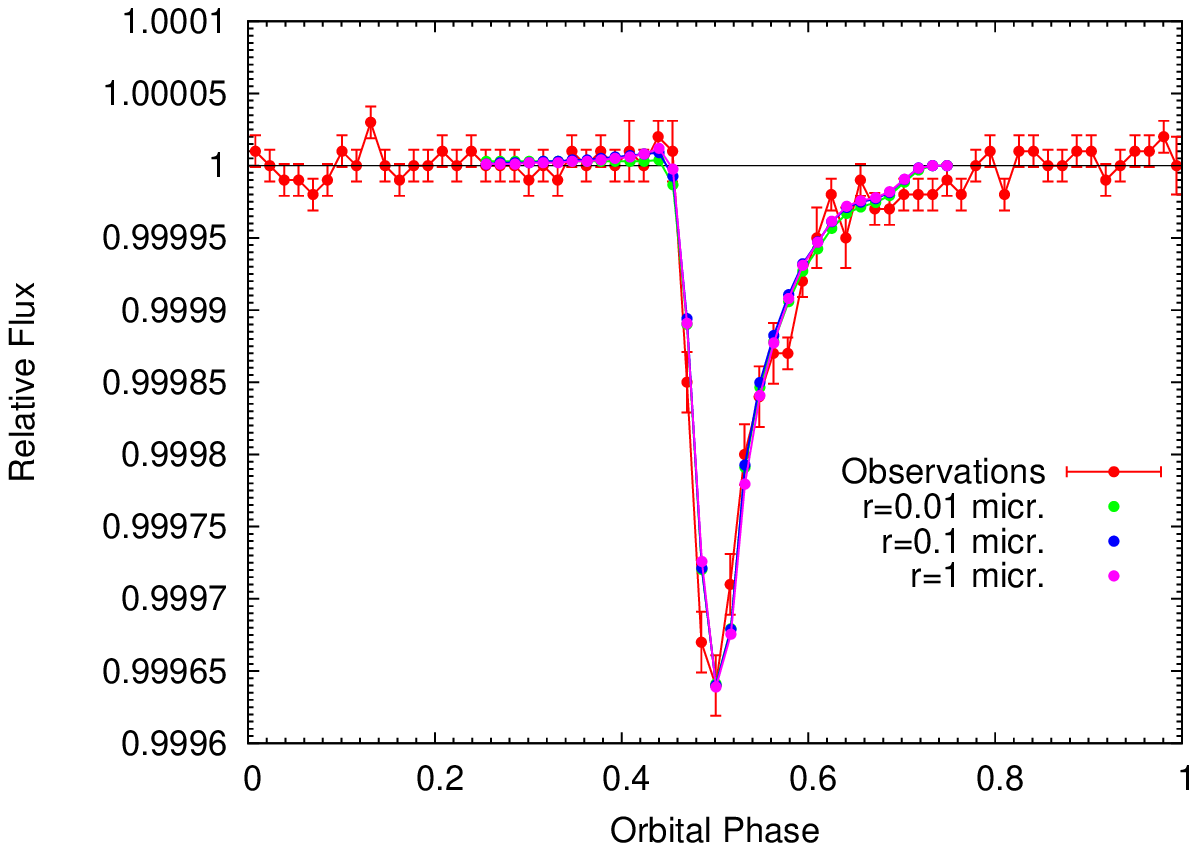}}
\caption{Model light curves calculated for 0.01-micron, 0.1-micron, and 1-micron grains of $\gamma$-alumina (left) and enstatite (right), compared with the observed light curve of KOI 2700b.}
\label{alumina_enstatite_a2m10}
\end{figure*}

\begin{figure*}
\centering
\centerline{
\includegraphics[width=80mm]{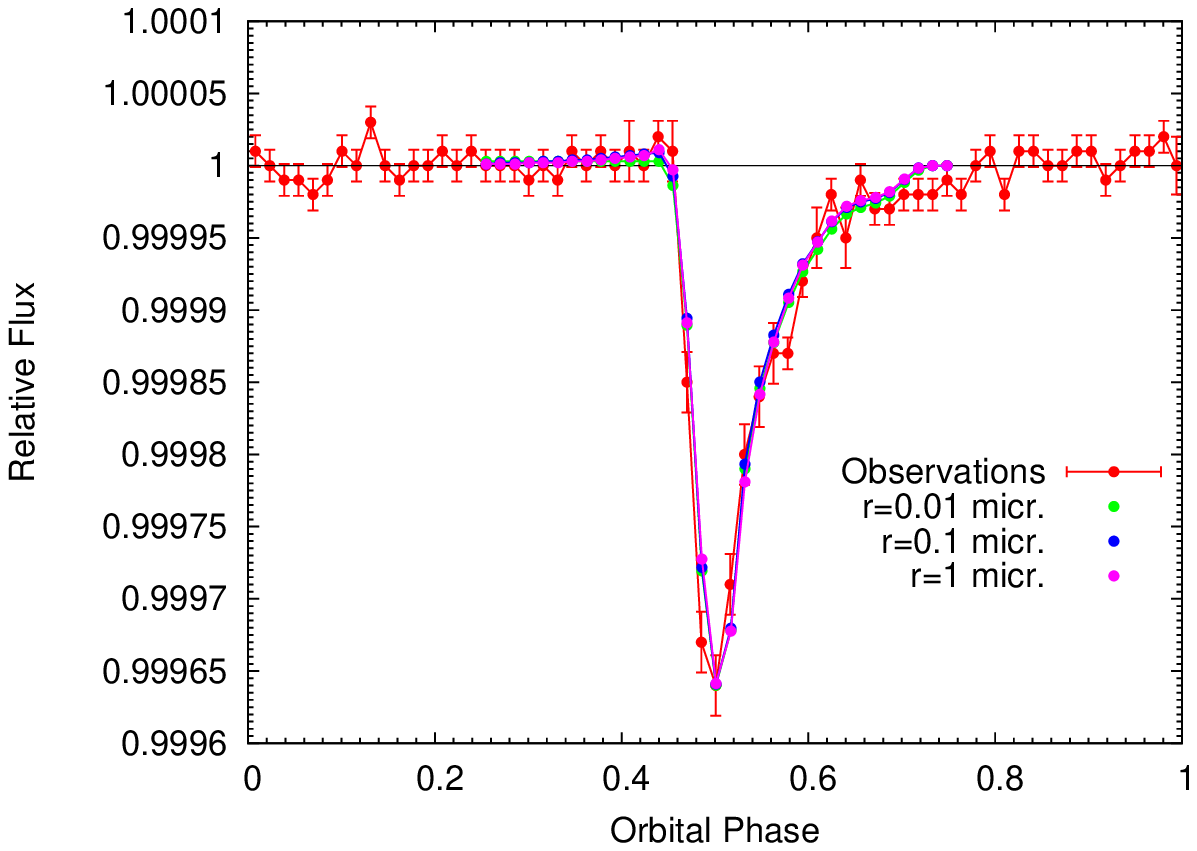}
\includegraphics[width=80mm]{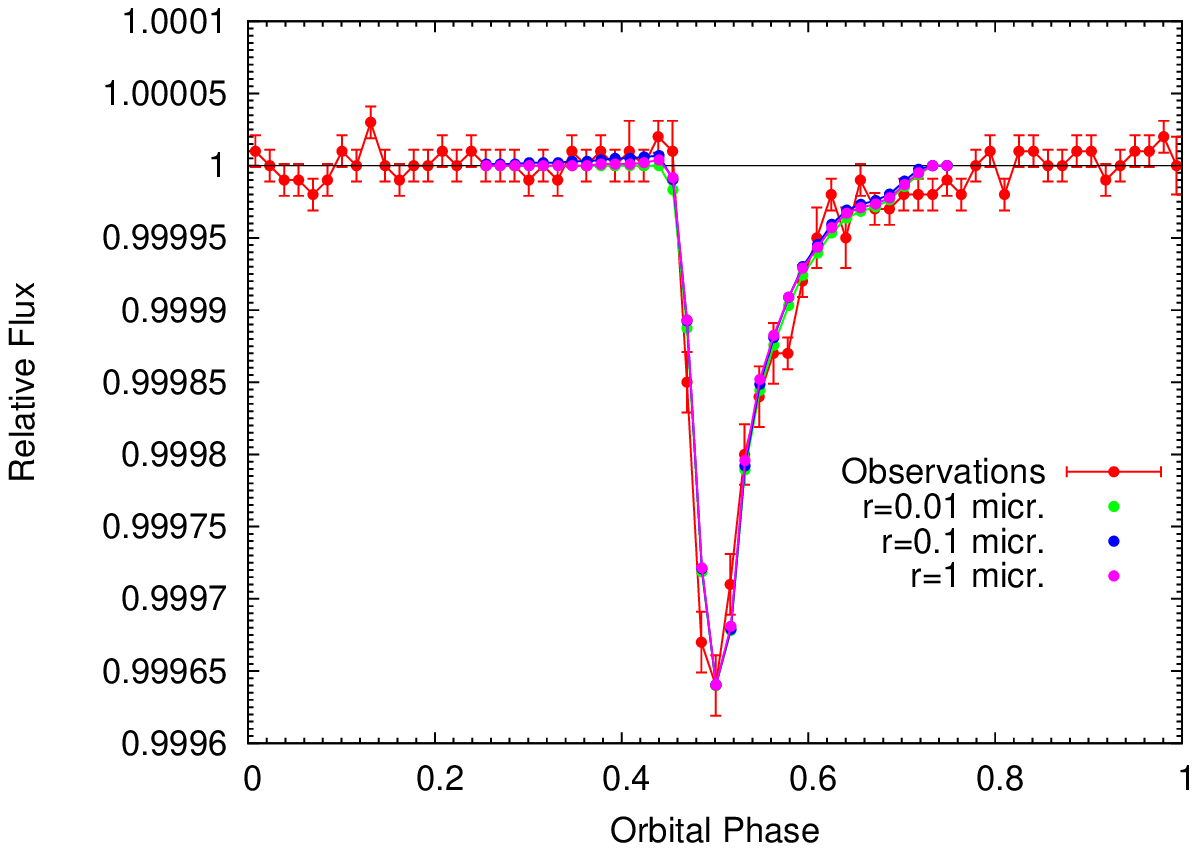}}
\caption{Model light curves calculated for 0.01-micron, 0.1-micron, and 1-micron grains of forsterite (left) and olivine (right), compared with the observed light curve of KOI 2700b.}
\label{forsterite_olivine_a2m10}
\end{figure*}

\subsection{Properties of the dust}
\label{dustres}
A decrease in dust density was detected in the comet-like tail of KIC 12557548b by \citet{Budaj1}. This feature of the tail is described by the density exponent $A1$ or $A2$ (see Eqs. \ref{densitychangeA1} and \ref{densitychangeA2}). As mentioned in Sect. \ref{modeling}, for our purposes we used the density exponent $A2$. It is one of the most important free parameters in our model. We first assumed that the dust density decreases along the tail of KOI 2700b in a similar manner to KIC 12557548b, with the exponent $-20$ or $-25$. Based on our fitting procedure, however, we obtained $A2$ in the range  $[-6.5,-8.2]$ (see Table \ref{partab} for individual values). The average value of the parameter is $A2=-7.13$. In the case of 1-micron $\gamma$-alumina grains the resulting parameter value was $A2\approx-8.0$, and in the cases of 1-micron enstatite grains and 1-micron forsterite grains we obtained $A2\approx-7.5$. For other dust particle sizes and species we obtained $A2\approx-7.0$ as a resulting parameter value. We note that the range of values found for the parameter $A2$ is in good agreement with the exponential decay factor value ($S$), found by \citet{Rappaport2}. Ignoring the $C(t)$ dependence of the dust density in the tail, the density exponent $A2$ can be converted to the exponential decay factor as $S=-A2/\pi$; hence the range of $A2$ values ($[-6.5,-8.2]$; see Table \ref{partab} for individual values) corresponds to $S$ of about $[2.1,2.6]$. The exponential decay factor value, found by \citet{Rappaport2}, is $S=2.4 \pm 1.0$.  

The density exponent $A2$ needs a certain density at the beginning of the ring $\rho(0)$ to fit the transit depth properly at each chemical composition and dust particle size. In Table \ref{partab} we present the resulting values for this free parameter and also the resulting values for the derived parameter $\tau_\mathrm{max}$, which is the highest optical depth through the tail in front of the star, and which follows from $\rho(0)$. These values were derived based on a tail model with a cross-section of $0.05 \times 0.05 R_\odot$ at the beginning and $0.09 \times 0.09 R_\odot$ at its end. Table \ref{partab} shows that 0.01-micron and 1-micron grains have very similar individual values of $\rho(0)$. On the other hand, the resulting values of $\tau_\mathrm{max}$ are different in the case of these two groups. The difference is about $3.00 \times 10^{-3}$. The third group, with 0.1-micron grains, has individual values of $\rho(0)$ that are  about one order of magnitude smaller, but similar resulting values of $\tau_\mathrm{max}$ to 0.01-micron grains. The average values of the parameters are $\rho(0)=0.1667 \times 10^{-15}$ g.cm$^{-3}$ and $\tau_\mathrm{max}=5.24 \times 10^{-3}$, and we obtained these parameters in the range $[0.0151,0.3333] \times 10^{-15}$ g.cm$^{-3}$ and $[4.28,7.43] \times 10^{-3}$, respectively.

\begin{figure}
\centering
\includegraphics[width=80mm]{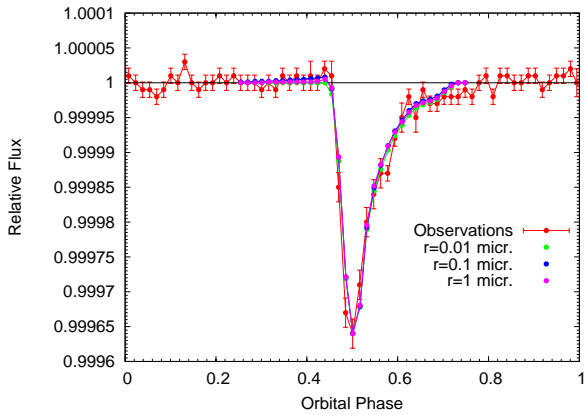}
\caption{As in Fig. \ref{forsterite_olivine_a2m10}, but for pyroxene grains.}
\label{pyroxene_a2m10}
\end{figure}

Based on modeling of the transit we cannot determine the chemical composition of the dust ejected by KOI 2700b. That is why we applied selected species, according to \citet{vanLieshout1}. Theoretically, the situation in the case of particle sizes is slightly better. We can estimate the typical particle size in the dust tail based on part of the light curve of KOI 2700b, where a pre-transit brightening is expected or possible, as was done at KIC 12557548b by \citet{Budaj1}. This brightening is caused by forward scattering on dust particles in the tail \citep{Brogi1, Budaj1} and the forward scattering is sensitive to the dust particle size (Fig. \ref{phasef}). Therefore, we can compare our model light curves with observations, especially in phases between 0.40 and 0.45, where the pre-transit brightening was detected at KIC 12557548b (see Fig. \ref{KICKOI}), and search for the particle size, which satisfies the observed light curve of KOI 2700b better. Figures \ref{alumina_enstatite_a2m10} -- \ref{pyroxene_a2m10} show that this part of the model light curve is sensitive to the dust particle size, however, differences between the model pre-transit brightenings are not enough to draw any satisfactory conclusions on the typical grain size in the dust tail. Since every model pre-transit brightening satisfies the observations, we cannot select a model with appropriate grain size.

\subsection{The dimension of the planet solid body}
\label{otherres}
\begin{figure}
\centering
\includegraphics[width=80mm]{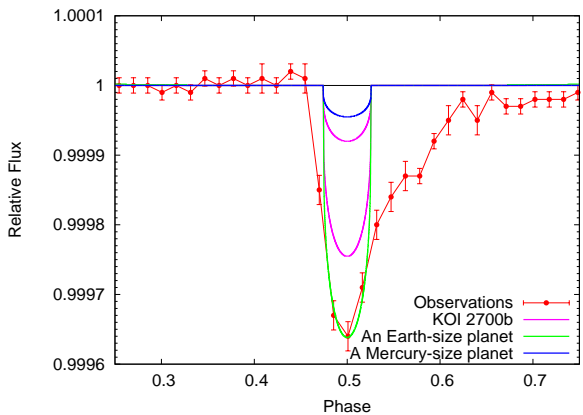}
\caption{Phase-folded and binned transit light curve of KOI 2700b (observations) compared with the transit light curve of the solid-body of KOI 2700b (according to the average value and upper limit of the planet radius; see the text). Furthermore, the transit light curve of an Earth-size planet and a Mercury-size planet is also plotted. The orbital inclination angle is $i=87$ deg.}
\label{earthmerk}
\end{figure}

The dimension of the solid body of KOI 2700b was a key point in our model. It was parametrized using the following free parameter:  the ratio of the radii ($R_\mathrm{p}/R_\mathrm{s}$). Based on the results of \citet{Rappaport2}, we selected the value of $R_\mathrm{p}/R_\mathrm{s}=0.017$ as an upper limit for this free parameter. However, as these authors pointed out, the corresponding value, $R_p=1.06R_{\oplus}$, needs to be improved. Therefore, we strongly expected that the planet radius would be smaller, $R_\mathrm{p}/R_\mathrm{s}\sim0.006$, which corresponds to a Mercury-sized planet. The parameter value $R_\mathrm{p}/R_\mathrm{s}\leq0.006$ is also predicted by the theory of the thermal wind and planet evaporation \citep{Perez1}. The resulting parameter values confirmed our expectations. Table \ref{partab} shows that in the cases of 1-micron $\gamma$-alumina, enstatite, and forsterite grains, we obtained $R_\mathrm{p}/R_\mathrm{s}\approx0.007$, in the cases of 0.01-micron enstatite and forsterite grains we obtained $R_\mathrm{p}/R_\mathrm{s}\approx0.008,$ and in other cases the resulting parameter value was 0.009. The average value of the parameter is $R_\mathrm{p}/R_\mathrm{s}=0.008$ ($R_\mathrm{p}=0.497R_\oplus$, or 3172 km). We obtained $R_\mathrm{p}/R_\mathrm{s}$ in the range $[0.0,0.014]$ (see Table \ref{partab} for individual values), hence we can consider the value of $R_\mathrm{p}/R_\mathrm{s}=0.014$, which corresponds to $R_\mathrm{p}=0.871R_\oplus$ (or 5551 km), as an upper limit on the radius of the planet solid body of KOI 2700b. The upper limit of the planet radius is in agreement with the theory of the thermal wind and planet evaporation and is also derived with higher accuracy than the result found by \citet{Rappaport2}. A comparison of the observed light curve with the solid-body transit light curve of KOI 2700b is depicted in Fig. \ref{earthmerk}. For a comparison, we also plotted the transit light curve of an Earth-size and a Mercury-size planet. The light curves were modeled using the software JKTEBOP\footnote{The software is available on the web-page \url{http://www.astro.keele.ac.uk/jkt/codes/jktebop.html}.} \citep{Southworth1} and its task No. 2. We applied $i=87$ deg (according to the average of listed values in Table \ref{partab}) and in the case of KOI 2700b the value of $R_\mathrm{p}/R_\mathrm{s}=0.008$ (the average value of the parameter) and $R_\mathrm{p}/R_\mathrm{s}=0.014$ (the upper limit of the planet radius).   

\subsection{Mass-loss rate from the planet}
We can estimate the mass-loss rate $\dot{M}$ from the planet assuming that the tail with the mass $M_\mathrm{tail}$ is replenished within a period, which is equal to the evaporation time scale of grains in the tail $T_\mathrm{evap}$:

\begin{equation}
\label{massloss0}
\dot{M}=\frac{M_\mathrm{tail}}{T_\mathrm{evap}}.
\end{equation}

The mass of the tail can be calculated by integration of Eq. \ref{densitychangeA2} through the volume of the tail $V$ as follows:

\begin{equation}
\label{massloss1}
M_\mathrm{tail}=\int\rho(0)\frac{C(0)}{C(t)}e^{(|t-t(0)|A2)/\pi}\mathrm{d}V.
\end{equation}

\noindent{Assuming that the geometrical cross-section of the tail is constant from the planet to the end of the ring, that is, $C(t)=C(0)$, and substituting $|t-t(0)|$ with $L/a$, where $L$ is the length of the tail and $a$ is the radius of the ring in cm, we can simplify Eq. \ref{massloss1} to:}

\begin{equation}
\label{massloss2}
M_\mathrm{tail}=\int\rho(0)e^{LA2/\pi a}\mathrm{d}V,
\end{equation}

\noindent{where the volume element of the tail can be expressed as $\mathrm{d}V=C(0)\mathrm{d}L$. After integration of Eq. \ref{massloss2} we get:}

\begin{equation}
\label{massloss3}
M_\mathrm{tail}=\rho(0)(e^{LA2/\pi a}-1)C(0){\frac{\pi a}{A2}}.
\end{equation}

The evaporation time scale of grains in the tail can be calculated using the equations and tables presented by \citet{vanLieshout1}. First, we can approximate the equilibrium vapor pressure as $p_\mathrm{v}=e^{(\mathcal{-A}/T_\mathrm{eqv}+\mathcal{B})}$ [dyn.cm$^{-2}$], where $T_\mathrm{eqv}$ is the equilibrium temperature of the dust grain and $\mathcal{A}$, $\mathcal{B}$ are material-dependent sublimation parameters. The values of $T_\mathrm{eqv}$ can be interpolated from the tables of \citet{Budaj3}, and the values of $\mathcal{A}$, $\mathcal{B}$ are tabulated for some species in \citet{vanLieshout1}. Subsequently, we can approximate the mass-loss flux from the surface of the dust grain $J$ [g.cm$^{-2}$.s$^{-1}$] as:

\begin{equation}
\label{massloss3.5}
J=\alpha p_\mathrm{v} \sqrt{\frac{\mu m_\mathrm{u}}{2\pi k_\mathrm{B} T_\mathrm{eqv}}},
\end{equation}

\noindent{where $\alpha$ is the evaporation coefficient, $\mu$ is the molecular weight of the molecules that sublimate, $m_\mathrm{u}$ is the atomic mass unit and $k_\mathrm{B}$ is the Boltzmann constant in cgs units. The values of $\alpha$ and $\mu$ are also tabulated for some species in \citet{vanLieshout1}. Finally, we can calculate the evaporation time scale of grains in the tail as:}

\begin{equation}
\label{massloss3.6}
T_\mathrm{evap} = \frac{r\rho_\mathrm{d}}{J},
\end{equation}

\noindent{where $r$ is the grain radius and $\rho_\mathrm{d}$ is the bulk density of dust species. The values of $\rho_\mathrm{d}$ are tabulated for some species in \citet{vanLieshout1} or in \citet{Budaj3}. Using Eqs. \ref{massloss3} and \ref{massloss3.6} we can rewrite Eq. \ref{massloss0} to the following final form:}

\begin{equation}
\label{massloss4}
\dot{M}=\frac{\rho(0)(e^{L A2/\pi a}-1)C(0){\frac{\pi a}{A2}}}{\frac{r\rho_\mathrm{d}}{J}}.
\end{equation}

For our purposes we used $C(t)=C(0)=0.05 \times 0.05 R_\odot$, the parameter values presented in Table \ref{partab} of this work, in Tables 2 and 3 of \citet{vanLieshout1} and in the Tables of \citet{Budaj3}. We made three groups of estimations of the mass-loss rate from the planet: one for alumina, one for enstatite, and one for forsterite. We note that in the cases of enstatite and forsterite we interpolated the equilibrium temperature of pyroxene with 40\% magnesium and 60\% iron, and olivine with 50\% magnesium and 50\% iron, respectively. Pure enstatite or forsterite have too low a value of $T_\mathrm{eqv}$, out of the temperature range for which $\mathcal{A}$ and $\mathcal{B}$ were determined. Our individual results are presented in Table \ref{masslosstab}. This procedure yields an average mass-loss rate from KOI 2700b of $\dot{M}=3.03 \times 10^{14}$ g.s$^{-1}$, and we obtained $\dot{M}$ in the range $[5.05 \times 10^{7},4.41 \times 10^{15}]$ g.s$^{-1}$. This range of values is consistent with the mass-loss rate estimate of $6 \times 10^9$ g.s$^{-1}$, presented by \citet{Rappaport2}.

\begin{table}
\caption{Estimation of the mass-loss rate from KOI 2700b.}
\centering
\begin{tabular}{llll}
\hline
\hline
Species         & $\dot{M}_\mathrm{min}$ [g.s$^{-1}$]   & $\dot{M}_\mathrm{med}$ [g.s$^{-1}$]    & $\dot{M}_\mathrm{max}$ [g.s$^{-1}$]\\ 
\hline
\multicolumn{4}{c}{$r=0.01$ micron}\\
Alumina         & $5.78 \times 10^{11}$                 &       $6.94 \times 10^{11}$            & $8.34 \times 10^{11}$ \\
Enstatite       & $1.39 \times 10^{15}$                 &       $2.48 \times 10^{15}$            & $4.41 \times 10^{15}$ \\
Forsterite      & $3.30 \times 10^{13}$                 &       $4.85 \times 10^{13}$            & $7.37 \times 10^{13}$ \\
\hline
\multicolumn{4}{c}{$r=0.1$ micron}\\
Alumina         & $1.10 \times 10^{11}$                 &       $1.11 \times 10^{11}$            & $1.12 \times 10^{11}$ \\
Enstatite       & $8.17 \times 10^{13}$                 &       $1.94 \times 10^{14}$            & $4.63 \times 10^{14}$ \\
Forsterite      & $9.31 \times 10^{11}$                 &       $1.55 \times 10^{12}$            & $2.59 \times 10^{12}$ \\
\hline
\multicolumn{4}{c}{$r=1$ micron}\\
Alumina         & $5.05 \times 10^{7}$                  &       $7.54 \times 10^{7}$         	 & $1.11 \times 10^{8}$  \\
Enstatite       & $7.01 \times 10^{10}$                 &       $7.83 \times 10^{10}$            & $8.85 \times 10^{10}$ \\
Forsterite      & $8.20 \times 10^{8}$                  &       $8.73 \times 10^{8}$         	 & $9.32 \times 10^{8}$  \\
\hline
Average value   & --                                    &       $3.03 \times 10^{14}$            & --                    \\
Extreme values  & $5.05 \times 10^{7}$                  &       --                               & $4.41 \times 10^{15}$ \\        
\hline
\hline                                          
\end{tabular}
\tablefoot{The Table contains the individual median values ($\dot{M}_\mathrm{med}$), as well as the individual minimum and maximum values of the mass-loss rate ($\dot{M}_\mathrm{min}$ and $\dot{M}_\mathrm{max}$), propagated from the uncertainties of contributing parameters. Also presented here is the average value of $\dot{M}_\mathrm{med}$. Extreme values are the lowest value of $\dot{M}_\mathrm{min}$ and the highest value of $\dot{M}_\mathrm{max}$.}     
\label{masslosstab}
\end{table}

\subsection{Comparison with KIC 12557548b}

Since the first exoplanet with a comet-like tail, KIC 12557548b, was analyzed similarly by \citet{Budaj1}, it is very interesting to compare his results with the results obtained in this work. Unfortunately, we cannot compare the typical grain size of the tail, which we could not derive in the case of KOI 2700b, and the dimension of the planet solid body, which was not included in the model of KIC 12557548b, used by \citet{Budaj1}. On the other hand, an upper limit on the radius of the planet solid body of KIC 12557548b was presented by \citet{Brogi1} and by \citet{Werkhoven1}, which we can use for our comparison. Similarly, the mass-loss rate from KIC 12557548b was calculated, for example, by \citet{Rappaport1}.

KOI 2700b and KIC 12557548b have very similar light-curve shape with some peculiarities. Mainly, the light curve displays a sharp ingress, followed by a short sharp egress and a long smooth egress. The light curve of KIC 12557548b shows a relatively significant pre-transit brightening as well, which in the light curve of KOI 2700b is not significant or even present at all. A relatively weak post-transit brightening is visible only in the light curve of KIC 12557548b. The transit duration in phase units is also very similar. The transits of the planets are very probably not edge-on. It seems that KOI 2700b has an orbital inclination angle of $i\approx87$ deg, while KIC 12557548b has $i\approx82$ deg. On the other hand, the transit depth of the planets is very different, KOI 2700b has an average transit depth about ten times shallower. Due to the similar light curves, the modified ring model of the tail fits the observations in both cases well. 

The tails are 60 deg long in both models, that is, in this work and in \citet{Budaj1}. The cut-off at this distance from the planet is an assumption of the model. The dust density in the tail decreases in both cases with the distance from the planet; however, these density decreasings are not similar. The dust density in the tail of KIC 12557548b decreases rapidly along the tail, which is well defined by the density exponent $A2\approx-20$ or $A2\approx-25$. This may produce the week, but still observable post-transit brightening at KIC 12557548b. In comparison with this planet, KOI 2700b exhibits a relatively low dust density decreasing in its tail, which is defined by the density exponent $A2\approx-7$. Therefore, in this case we cannot expect any post-transit brightening. 

KOI 2700b and KIC 12557548b may have a planet solid body radius less than $\sim 1 R_\oplus$. Since we obtained $R_\mathrm{p}/R_\mathrm{s}$ in the range $[0.0,0.014]$ (see Table \ref{partab} for individual values), we can consider the value of $R_\mathrm{p}/R_\mathrm{s}=0.014$, which corresponds to $R_\mathrm{p}=0.871 R_\oplus$, or 5551 km, as an upper limit of the planet radius of KOI 2700b. In comparison with KOI 2700b, KIC 12557548b may have a slightly bigger radius as well, up to $\sim 1 R_\oplus$. \citet{Brogi1} present the value of $R_\mathrm{p}=1.15 R_\oplus$ and \citet{Werkhoven1} the value of $R_\mathrm{p}=0.72 R_\oplus$ as an upper limit for the radius of the planet solid body of KIC 12557548b, which corresponds to about 7300 km and 4600 km, respectively.

Although both planets are disintegrating and losing material, neither KOI 2700b nor KIC 12557548b  show any evidence of a significant long-term orbital period variability. This is in agreement with the prediction, presented by \citet{Perez1}, which suggests that disintegrating planets undergo negligible orbital period variations due to the evaporation process. A significant variability is observable in the light-curve shape of the planets. In the case of KIC 12557548b there is a short-term periodic variability in the transit core \citep{Rappaport1} and a long-term periodic variability in the egress \citep{Budaj1}. KOI 2700b exhibits only a long-term variability in the transit core, which may be periodic as well.    

Based on our estimations, the mass-loss rate from KOI 2700b may be from $5.05 \times 10^{7}$ to $4.41 \times 10^{15}$ g.s$^{-1}$. This interval of values is consistent with the upper limit of the mass-loss rate from KIC 12557548b, which was estimated by \citet{Rappaport1}, and which is $2 \times 10^{11}$ g.s$^{-1}$.

\section{Conclusions}
\label{cnc}

Our main scientific goal was to verify the disintegrating-planet scenario of KOI 2700b by modeling its light curve and put constraints on various tail and planet properties. The orbital period of the planet was improved based on the \textit{Kepler} data from quarters 1 -- 17. We obtained the value of $P_\mathrm{orb}=0.9100259(15)$ days, which is in good agreement with the orbital period presented by the discoverers. We also searched for long-term orbital period changes of KOI 2700b, however, in agreement with \citet{Perez1}, we found no significant evidence. We can confirm the decrease in transit depth of the exoplanet found by \citet{Rappaport2}. On the other hand, we cannot exclude the possibility that the transit depth varies periodically over a long-term time span longer than the available \textit{Kepler} quarters.

We obtained the phase-folded and binned transit light curve of KOI 2700b, which we iteratively modeled using the radiative-transfer code SHELLSPEC. Mie absorption and scattering on spherical dust grains with realistic dust opacities, phase functions, and finite radius of the source of the scattered light were taken into account. During the modeling we applied selected species ($\gamma$-alumina, enstatite, forsterite, olivine with 50\% magnesium and 50\% iron and pyroxene with 40\% magnesium and 60\% iron) and dust particle sizes (0.01, 0.1 and 1 micron in radii). We executed 15 joint fitting procedures -- one for each combination of species and dust particle size. We modeled the comet-like tail as part of a ring around the parent star. The geometrical cross-section of the ring is monotonically enlarging from the planet to the end of the ring. Since this modified ring model of the tail satisfies the observations well, we confirm the disintegrating-planet scenario of KOI 2700b. 

Furthermore, via modeling, we derived some interesting features of KOI 2700b and its comet-like tail. We first assumed that the orbital plane of the exoplanet and its tail had an inclination of $i=90$ deg with respect to the plane of the sky, however it turns out that the orbital plane of the planet is not edge-on, but the orbital inclination angle is very probably close to the value of $i=87$ deg. We obtained $i$ in the range $[85.1,88.6]$ deg. KOI 2700b exhibits a relatively low dust density decreasing in its tail, which is defined by the density exponent $A2 \approx -7$. During our modeling process, we obtained this parameter in the range $[-6.5,-8.2]$, which is in good agreement with \citet{Rappaport2}. We also derived the dust density at the beginning of the ring ($\rho(0)$) and the highest optical depth through the tail in front of the star ($\tau_\mathrm{max}$), based on a tail model with a cross-section of $0.05 \times 0.05 R_\odot$ at the beginning and $0.09 \times 0.09 R_\odot$ at its end. The average values of the parameters are $\rho(0)=0.1667 \times 10^{-15}$ g.cm$^{-3}$ and $\tau_\mathrm{max}=5.24 \times 10^{-3}$, and we obtained these parameters in the range $[0.0151,0.3333] \times 10^{-15}$ g.cm$^{-3}$ and $[4.28,7.43] \times 10^{-3}$, respectively. Since there is still ample room for a contribution to the light curve from the solid body of the planet, we also included it in the model. Our average value for the dimension of the planet is $R_\mathrm{p}/R_\mathrm{s}=0.008$ ($R_\mathrm{p}=0.497 R_\oplus$, or 3172 km). Since we obtained this parameter in the range $[0.0,0.014]$, we can consider the value of $R_\mathrm{p}/R_\mathrm{s}=0.014$, which corresponds to $R_\mathrm{p}=0.871R_\oplus$ (or 5551 km), as an upper limit on the radius of the planet solid body of KOI 2700b. The upper limit of the planet radius is in agreement with the theory of the thermal wind and planet evaporation \citep{Perez1} and is also derived with higher accuracy than the result found by \citet{Rappaport2}. We estimated the mass-loss rate from KOI 2700b, and we obtained $\dot{M}$ values from the interval $[5.05 \times 10^{7}, 4.41 \times 10^{15}]$ g.s$^{-1}$, consistent with the estimate of \citet{Rappaport2}. Our average estimated mass-loss rate from KOI 2700b is $\dot{M}=3.03 \times 10^{14}$ g.s$^{-1}$. On the other hand, we could not draw any satisfactory conclusions on the typical grain size in the dust tail due to the small differences between the model pre-transit brightenings at different particle radii. Finally we showed that the first exoplanet with a comet-like tail, KIC 12557548b, and KOI 2700b are similar in many respects.  

\begin{acknowledgements}
I thank Dr. J. Budaj for the technical assistance, comments, and discussions. I also thank the anonymous referee for helpful comments and corrections. This work was supported by the Slovak Central Observatory Hurbanovo, by the VEGA grant of the Slovak Academy of Sciences No. 2/0031/18 and by the realization of the Project ITMS No. 26220120029, based on the Supporting Operational Research and Development Program financed from the European Regional Development Fund. 
\end{acknowledgements}

\bibliographystyle{aa} 
\bibliography{Yourfile} 

\end{document}